\documentclass[sigconf]{acmart}
\AtBeginDocument{%
  }

\setcopyright{acmlicensed}
\copyrightyear{2018}
\acmYear{2018}
\acmDOI{XXXXXXX.XXXXXXX}
\acmConference[Conference acronym 'XX]{Make sure to enter the correct
  conference title from your rights confirmation email}{June 03--05,
  2018}{Woodstock, NY}

\newcommand{\one}{({\em i}\/)\xspace}
\newcommand{\two}{({\em ii}\/)\xspace}
\newcommand{\three}{({\em iii}\/)\xspace}

\newcommand{\pb}[1]{\vspace{0.75ex}\noindent{\bf \em #1}}
\usepackage{tablefootnote}
\usepackage{xspace}
\usepackage{csquotes}
\usepackage{tcolorbox}
\usepackage{subcaption}
\usepackage{soul}
\usepackage{makecell}
\usepackage{mwe}
\usepackage{multirow}
\usepackage{graphicx}
\usepackage{nicematrix}
\usepackage{fontawesome}
\usepackage{tabularx}
\usepackage{booktabs}
\usepackage{pifont}
\usepackage{multirow}
\usepackage{amsmath}
\usepackage{svg}
\usepackage{verbatim}
\usepackage{xcolor}
\usepackage{xstring}
\usepackage{enumitem}
\usepackage{array}
\usepackage{diagbox}
\usepackage{textcomp}
\usepackage{bm}
\newcolumntype{L}[1]{>{\raggedright\let\newline\\\arraybackslash\hspace{0pt}}m{#1}}
\newcolumntype{C}[1]{>{\centering\let\newline\\\arraybackslash\hspace{0pt}}m{#1}}
\newcolumntype{R}[1]{>{\raggedleft\let\newline\\\arraybackslash\hspace{0pt}}m{#1}}
\newcommand\rot[1]{\rotatebox[origin=c]{90}{#1}}
\newcommand\Tr[1]{\begin{tabular}[t]{@{}c@{}}#1\end{tabular}}
\begin{document}

\title{The Invisible Hand: Characterizing Generative AI Adoption and its Effects on An Online Freelancing Market}

\author{Yiming Zhu}
\authornote{Yiming Zhu is also with HKUST (GZ), Gareth Tyson is also with Queen Mary University of London, and Pan Hui is also with the University of Helsinki.}
\affiliation{%
  \institution{HKUST}
 \city{Hong Kong SAR}
 \country{China}
}

\author{Gareth Tyson}
\affiliation{%
  \institution{HKUST (GZ)}
  \city{Guangzhou}
  \country{China}
}

\author{Pan Hui}
\affiliation{%
  \institution{HKUST (GZ)}
  \city{Guangzhou}
  \country{China}
}



\begin{abstract}
 Since the COVID-19 pandemic, freelancing platforms have experienced significant growth in both worker registrations and job postings. However, the rise of generative AI (GenAI) technologies has raised questions about how it affect the job posting in freelancer market. Despite growing discussions, there is limited empirical research on the GenAI adoption and its effect on job demand and worker engagement. We present a large-scale analysis of Freelancer.com, utilizing over 1.8 million job posts and 3.8 million users. We investigate the emergence of jobs with the adoption of GenAI and identify leading position of ChatGPT in the freelancing market. With a focus on ChatGPT related jobs, we inspect their specific skill requirements, and the tasks that workers are asked to perform. Our findings provide insights into the evolving landscape of freelancing in the age of AI, offering a comprehensive profile of GenAI's effects on employment, skills, and user behaviors in freelancing market.
\end{abstract}

\ccsdesc[500]{Human-centered computing~Empirical studies in collaborative and social computing}

\keywords{Quantitative methods, Empirical analysis, Online Labor Market, Freelancer.com, Generative AI, ChatGPT}


\maketitle

\section{Introduction}\label{sec:intro}

Platforms like Freelancer.com, Fiverr, and Upwork have emerged as important marketplaces where individuals or companies can hire freelance workers --- independent professionals offering a wide range of services on a project-by-project basis. These platforms are central to the gig economy, facilitating flexible work arrangements by connecting employers with Freelance workers, who possess the desired skills. Since the onset of the COVID-19 lockdowns, these platforms have experienced a surge in new registrations and job postings~\cite{gig-is-up, munoz2022new}. Today, freelancing platforms attract millions of users, and have become a vital source of employment in the digital economy~\cite{freelancer-report2}.

Recently, discussions about the role of generative AI (GenAI) on the labor market have intensified, due to their ability to handle repetitive and computation-heavy work~\cite{10.1145/3589335.3641295, KSHETRI2024102716}. Thus, some argue that GenAI poses an existential threat to jobs in many fields such as content writing, design, and human resources~\cite{lazaroiu2023generative, yilmaz2023ai}. 
Indeed, anecdotal evidence has suggested that GenAI has already led to a decline in demand for jobs prone to automation and manual-intensive work~\cite{demirci2023ai}.
Freelance bidders with skills in writing and graphic design may therefore face increased competition following the rise of GenAI~\cite{gen-ai-disrupt-creative}. 
Moreover, it is reported that users with suitable expertise are increasingly focusing their efforts \emph{exclusively} on GenAI related jobs~\cite{liu2023generate, yiu2024ai}. Despite the growth of these trends, we lack any empirical study profiling the effect of GenAI on employment trends, job content, and employee behaviors.

To address this deficit, we perform a large-scale empirical study of the role of 
GenAI in online freelancing markets. To study this, we gather data from Freelancer.com, one of the largest global freelancing platforms for job posting. Freelancer.com has over 64 million registered users and 22.2 million job posts~\cite{freelancer-report} (2023). Such a large scale makes it the perfect candidate to explore how jobs related to GenAI emerge and their effects on the existing freelancing market. For this, we compile a two-year dataset containing all metadata for 1.8M+ job posts and 3.8M+ users since $1^{st}$, January, 2022. Using this data, we explore the following research questions:

\begin{itemize}[leftmargin=*]

    \item \textbf{RQ1:} How often are GenAI technologies mentioned in job posts? Do these jobs attract larger budgets or more bids from the employers and bidders? Which GenAI technologies are most commonly required by such jobs?

    \item \textbf{RQ2:} What types of employers and bidders (job applicants) engage in GenAI-related freelance jobs? Are the employers and bidders who use GenAI ``isolated'' from those who have do not use GenAI?

    \item \textbf{RQ3:} What uses of GenAI are requested in freelance jobs? Do these uses affect other user behaviors, e.g., job posting frequency?

    \item \textbf{RQ4:} Given the interest in understanding which jobs will remain relevant in the era of GenAI, can we predict which job skills are sought by jobs related to GenAI?
    
\end{itemize}

In answering the above questions, we provide a large-scale empirical analysis for GenAI's impact on the freelancing market. As a key time point, we focus on the changes after ChatGPT's release date to study how these technologies affect job patterns. 
Our main findings include:

\begin{enumerate}[leftmargin=*]

    \item After ChatGPT's release, GenAI-related jobs become more popular than non-GenAI jobs.
    Our causal effect analysis reflects that ChatGPT's release does play a role in catalyzing more demand on GenAI. Meanwhile, ChatGPT present a leading position in the GenAI freelancing market. 29.48\% skills and 32.05\% job posts related to GenAI mention ChatGPT. ChatGPT possesses $1.7\times$ the job post volume, as many as the total job post volume of other GenAI (e.g., DALL-E, Copilot, Gemini, Llama, ...).
    
    \item There are also clear price differentials. Jobs involving GenAI are relatively higher paying ($1.42\times$ median value of minimum budget of non-GenAI-related jobs) and attract more bids from bidders (bid count $1.88\times$, bid price $1.67\times$). We posit this is because GenAI-related jobs tend to derive from competitive industries, where jobs receive more bids.

    \item Adopting GenAI has become common among leading users with good reputations. Such users tend to possess central positions in the bid network.
    For example, compared to users who haven't released or bid on any GPT-related jobs, they receive higher ratings for their overall performance (employers: $1.01\times$; bidders: $5.14\times$) and possess higher PageRank centrality (employers: $9.89\times$; bidders: $4.18\times$) within the bid network.
    Moreover, users engaged in GenAI-related jobs are not isolated from those who are not. After ChatGPT's release, the division between them and those without experience in GPT-related jobs is shrinking on the bid network. We see a decreasing network modularity and an increasing proportion of inter-community edges. These indicates users (employers and bidders) of GenAI have been embraced by the wider community of bidders.
\end{enumerate}
Given the leading role of ChatGPT on Freelancer.com, we focus on it as a case study to provide a concrete lens to explore how employers and bidders adopt GenAI in freelance jobs for answering RQ3 and RQ4. Our findings include:

\begin{enumerate}[resume, leftmargin=*]
    \item We uncover two main uses of ChatGPT:
    integrating ChatGPT into software (covering 49.28\% GPT-related job posts) and reproducing human-generated content (26.69\%). We further profile the role of ChatGPT in shifting behaviors. Employers who post jobs that involve ChatGPT, tend to continue using it: 28.99\% of employers' jobs, released after their first ChatGPT use, also require ChatGPT. 
    
    \item Bidders expand their listed job skills after first bidding for GPT-related jobs. They engage in jobs that mention a wider range of job skills from more diverse industries, with their skills count rising from 59.0 to 66 (median).

    \item We uncover nine features that can effectively predict which job skills will be relevant to ChatGPT jobs. These features only rely on market data before ChatGPT's release. Classifiers using these features can accurately predict the skills growing in prominence for ChatGPT jobs after ChatGPT's release, achieving 86.81\% accuracy with 72.91\% F1-score.

\end{enumerate}


\section{Background \& Data Collection}\label{sec:Method}

We utilize Freelancer.com's API to compile a dataset containing all job posts, bids, job skills, and users on Freelancer.com, between {1$^{st}$, January, 2022} to {21$^{st}$, April, 2024}.

\pb{Primer on Freelancer.com.} 
Freelancer.com is an online freelancing marketplace founded in 2009. It allows businesses to release job posts and hire independent professionals. Users on Freelancer.com involve \textit{employers} and workers, who are referred to as \textit{bidders}. Employers release \textit{job posts} with their \textit{budgets} on the platform and bidders submit \textit{bids} to apply for a particular job. A bid contains the bidder's proposal for the work and the expected payment to be received. The employer then manually selects which bidder to assign the job to.

\pb{Job post and bid data.} We collect 1,831,788 job posts and their 52,243,610 bids. Note, we later study changes that take place after ChatGPT's release on 30$^{th}$, November, 2022. 959,591 (52.39\%) of these job posts were posted after this date. For job posts, we collect their submission time, textual work descriptions, required job skills, the minimum/maximum budgets, the count and average price of bids received from other bidders. Note, the workers (bidders) must submit bids to any job they wish to work on, including the price they expect to be paid (i.e., their bid). Thus, we collect data for all bids, covering their price (in USD), and the submission time. 

\pb{Job skill data.} We collect 2,874 job skills from the job posts. A required skill is a specific tag (formally defined by Freelancer.com) that describes what skills are required by job posts, e.g., python programming, copy editing. 230 (8.00\%) of the skills first emerge after ChatGPT's release. We also collect the industry category the job skill belongs to, as defined by Freelancer.com.

\pb{User data.} We extract the 3,893,131 users, who have either posted a job or bid on a job at least once during our measurement period.
We collect their work reputation (work quality, level of professionalism, ratings of willingness to hire/work for again) and hourly rate.

\pb{Ethics consideration.} Our study utilizes publicly available data from the Freelancer.com API. We make no attempt to de-anonymize users, and our sole focus is to examine GenAI's impact. All analyses are conducted in accordance with Freelancer.com's data policies.


\section{Studying The Rise of GenAI Jobs}
\label{sec:RQ1}

\begin{figure*}[t]
    \centering
    \begin{subfigure}[b]{.24\linewidth}
        \centering
        \includegraphics[width=\linewidth]{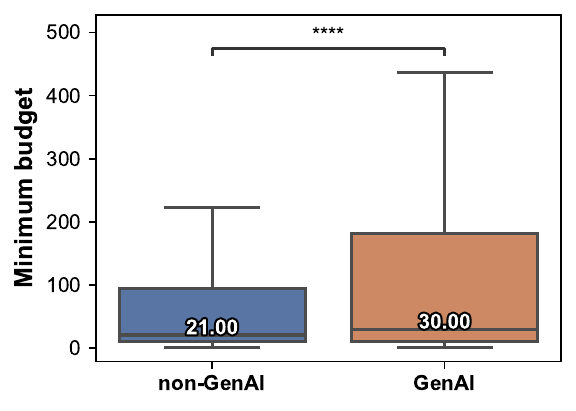} 
    \end{subfigure}
    \begin{subfigure}[b]{.24\linewidth}
        \centering
        \includegraphics[width=\linewidth]{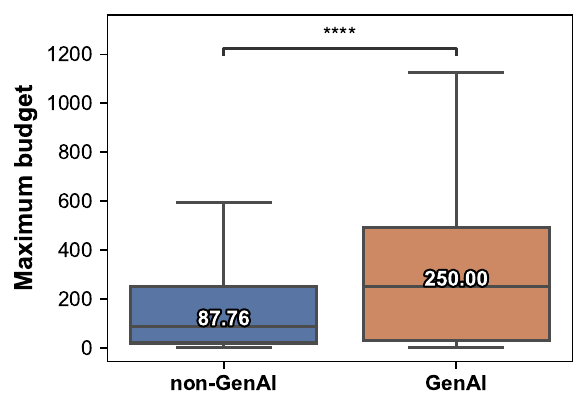}  
    \end{subfigure}
    \begin{subfigure}[b]{.24\linewidth}
        \centering
        \includegraphics[width=\linewidth]{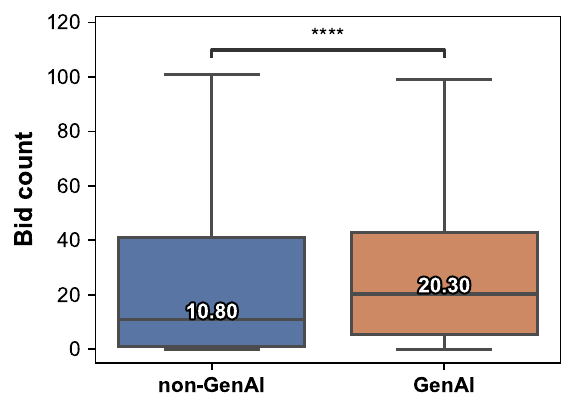}  
    \end{subfigure}
    \begin{subfigure}[b]{.24\linewidth}
        \centering
        \includegraphics[width=\linewidth]{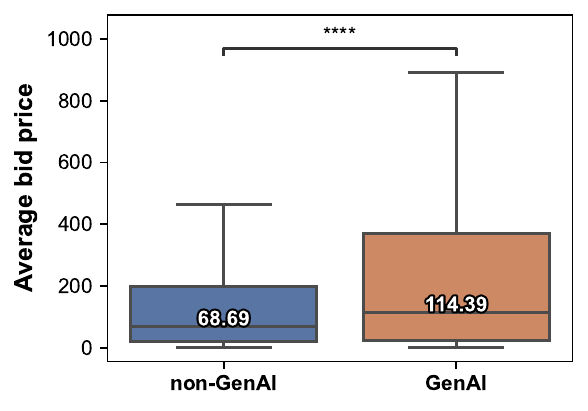} 
    \end{subfigure}
    \caption{Comparison of minimum/maximum budget, bid count, and average bid price between GenAI-related job posts and non-GenAI-related job posts. The statistical significance is reported by the Mann-Whitney U test; ****: $\bm{p<0.001}$.}
    \label{fig:post_compare}
\end{figure*}

We begin by exploring the emerging job skills and posts after ChatGPT's release.
We use this as the epoch, as this was arguably the time point where GenAI reached mainstream attention. We aim to study whether jobs related to GenAI now possess greater popularity than non-GenAI job posts (\textbf{RQ1}).

\pb{GenAI skill are becoming more in demand.}
We first measure the number of new job skills that emerge after ChatGPT's release. Overall, there are 229 new job skills, mentioned across 6,395 job posts. Recall, a skill is a specific tag that describes what skills are required by job posts (e.g., programming).
Through manual inspection, we find that 78 new skills (33.91\%) are directly oriented towards GenAI.
In total, 3,527 of the job postings mentioning these new skills (55.15\%) require GenAI job skills. 
On average, GenAI related ($\mu=32.15$) job skills are required by more job posts than non-GenAI job skills ($\mu=23.67$).
These results show that, after ChatGPT's release, GenAI play a role in job posts that mention new job skills.

\pb{GenAI jobs are becoming more plentiful.}
Next, we examine the number of job posts associated with GenAI. 
To achieve this, we compile a taxonomy of GenAI related terms (Table~\ref{tab:glossary}). This is based on a GenAI glossary by Civitai, one of the largest GenAI platforms~\cite{genai-glossary, wei2024exploring}, plus all version names of popular GenAI models/applications: ChatGPT, Llama, DALL-E, Copilot, Gemini, Sora, MidJourney, Stable Diffusion, and Claude. Accordingly, we categorize job posts released after ChatGPT's release into two groups: 

\begin{itemize}[leftmargin=*]
    \item \textbf{\texttt{GenAI}}: Job posts containing any GenAI skills or keywords, but excluding phrases forbidding GenAI usage: ``no/none/not use \{any keywords in our taxonomy\}''.
    
    \item \textbf{\texttt{non-GenAI}}: Job posts not belonging to \texttt{GenAI} group.
\end{itemize}

Overall, 959,591 job posts are published after ChatGPT's release. We identify 16,277 (1.70\%) \texttt{GenAI} job posts. Only 135 (0.014\%) job posts are filtered out as forbidding GenAI usage. 
This indicates an emerging adoption of GenAI among employers in freelancing markets.
1.70\% of all new job posts involve GenAI work.

To study their growth pattern, we next inspect the temporal trends of \texttt{GenAI} vs. \texttt{non-GenAI} job posts. For each job post group, we perform a linear regression analysis to investigate how the daily job post count evolve across the days after ChatGPT's release. We emphasize that each model above has a different release time, yet we rely on ChatGPT's release date because this is the point where GenAI reached mainstream attention~\cite{gpt-kick-genai}.
We formulate this regression as $Daily~job~post~count=Trend*Days+Intercept+\epsilon$.
We find that \texttt{non-GenAI} job posts exhibit a negative trend ($Trend=-1.178, p<0.001$). In contrast, \texttt{GenAI} ($Trend=0.0295, p<0.001$) job posts display positive trends after the release of ChatGPT. These findings suggest that, unlike non-GenAI jobs which shows a decreasing popularity of job postings, jobs involving GenAI are becoming more popular in freelancing market following ChatGPT's release.

\begin{table}[t]
\resizebox{.9\linewidth}{!}{%
\begin{tabular}{|l|ccccc|}
\toprule
\textbf{Term} & \textbf{Coefficient} & \textbf{Std error} & \textbf{p-value} & \textbf{2.5\%} & \textbf{97.5\%}\\ \midrule
Intercept & 2717.270 & 28.408 & *** & 2661.591 & 2772.948 \\ 
$\text{AI}_i$ & -2580.865 & 25.807 & *** & -2631.447 & -2530.284 \\ 
$\text{Post}_t$ & -429.635 & 36.363 & *** & -500.905 & -358.364 \\ 
$\text{AI}_i\times\text{Post}_t$ & 749.537 & 31.258 & *** & 688.271 & 810.802 \\ 
$\text{TimeTrend}_t$ & -0.728 & 0.058 & *** & -0.841 & -0.615 \\ \midrule
{Adj. R-squared}
 & \multicolumn{5}{c|}{0.929} \\ 
{F-statistic}
 & \multicolumn{5}{c|}{5274.462} \\ 
{Prob (F-statistic)}
 & \multicolumn{5}{c|}{***} \\ 
 \bottomrule
\end{tabular}%
}
\caption{A summary of DiD model's result for examining the causal effect of ChatGPT's release on daily post count for GenAI jobs. ***: $p<0.001$.}
\label{tab:did}
\end{table}

\begin{figure}[t]
  \centering
  \includegraphics[width=.7\linewidth]{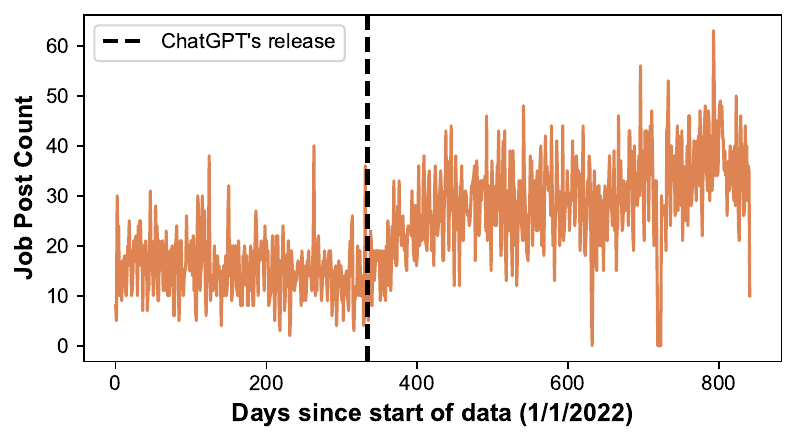}
  \caption{Temporal distribution of GenAI job post count.}
  \label{fig:did_genai_job}
\end{figure}

\pb{ChatGPT's release catalysts GenAI job demand.}
One intuitive explanation for the increasing trend of GenAI-related jobs is that the release of ChatGPT prompts more demand in GenAI related fields.
To explore this, we employ a Difference-in-Differences (DiD) design to estimate the causal effect of ChatGPT's release (30$^{th}$, November, 2022) on the demand for GenAI-related jobs, using non-GenAI jobs as a control group. The DiD approach compares the pre-post change in daily GenAI-related job posts count to the corresponding change in non-GenAI job posts count. We formulate our DiD models as:
\begin{equation}
\begin{aligned}
Y_{it} ={} & \beta_{0}+\beta_{1}\text{AI}_{i}+\beta_{2}\text{Post}_{t}+\beta_{3}(\text{AI}_{i}\times\text{Post}_{t})+\\
&\gamma\text{TimeTrend}_{t}+\epsilon_{it}
\end{aligned}
\end{equation}

\noindent where $Y_{it}$ denotes the Job post count for job category $i$ (GenAI or non-GenAI) at time $t$ (day). $\text{AI}_{i}$ is a binary indicator showing if the job category $i$ is GenAI (=1) or non-GenAI. $\text{Post}_{t}$ is a binary indicator showing if the current time $t$ is before ChatGPT's release (=1) or not. The intersection term $\text{AI}_{i}\times\text{Post}_{t}$ is the estimator capturing the causal effect of ChatGPT's release on job demand. $\text{TimeTrend}_{t}$ represents the continuous linear time trend, examined as days since start of data. $\epsilon_{it}$ is the error term.

Table~\ref{tab:did} summarizes the DiD model results examining the causal effect. The model demonstrates good explanatory power with an adjusted R-squared of 0.929, indicating that approximately 92.9\% of the outcome variation is explained by the specified model. 
We find that the effect estimator is associated with a positive coefficient with statistical significance ($\beta_3=749.537,p<0.001$). This suggests ChatGPT's release has a substantial positive effect, increasing the daily post count of GenAI-related jobs. 

For visualization, Figure~\ref{fig:did_genai_job} presents the temporal distribution of daily post count for GenAI jobs. Compared before ChatGPT's release, the trend of GenAI job posts presents a notable increase, with the average daily job post count increased by 88.93\% ($\mu_{\text{before}}=15.251, \mu_{\text{after}}=28.815$). Our findings suggest that the release of ChatGPT has a positive causal effect on the frequency of GenAI-related jobs on Freelancer.com, acting as a major catalyst for freelancing market demand in GenAI-related fields.

\pb{Fields related to GenAI.}
We next take a closer look on the fields associated with the increasing GenAI-related jobs.
Here, we assess the distribution of industries (as defined by the Freelancer.com API) among \texttt{GenAI} job posts.
Figure~\ref{fig:genai_industry} plots the number of Gen-AI related jobs posts tagged within each industry. We see that they have covered job skills from 17 diverse industries (note, a freelance job can exist in multiple industries). 
The most common industries are ``IT, Software'' (69.46\% job posts), ``Design, Media'' (36.58\%), ``Engineering'' (18.16\%), ``Writing'' (14.22\%), and ``Sales'' (10.36\%).
These observations also confirm recent economic studies~\cite{eloundou2023gpts, chui2023economic} that imply popular areas like development, writing, design, and sales of digital market may be heavily penetrated by GenAI.

\pb{GenAI-related jobs are more competitive.} 
The rise of GenAI-related jobs leads us to conjecture that these jobs may have different budgets and subsequently receive more bids from worker bidders. Thus, we next examine whether GenAI job posts differ from non-GenAI ones. Figure~\ref{fig:post_compare} presents the distributions of four metrics: minimum budget, maximum budget, bid count and average bid price (i.e., the bidders' expected payment for the jobs). Using the Mann-Whitney U (MWU) test, we find that \texttt{non-GenAI} job posts have significantly ($p<0.001$) \textit{lower} distributions than \texttt{GenAI} job posts across all four metrics. This indicates that employers tend to offer higher budgets for GenAI-related work (1.42× median value of
minimum budget of \textit{non-GenAI} jobs), and these job posts receive higher bid demands from bidders (1.67× bid prices in USD). Therefore, our results reveal GenAI's competitive edge in the freelancing market, as GenAI-related job posts often offer better salaries and attract more applications.

\pb{ChatGPT is leading the market.} 
We are also curious to explore which specific GenAI technologies are leading the freelancing market. Thus, we identify nine GenAI models reported as commonly used: ChatGPT, DALL-E, Copilot, Gemini, Sora, Llama, MidJourney, Stable Diffusion, and Claude. For each of these GenAI models, we manually create a list of relevant job skills and keywords (Table~\ref{tab:glossary}). Accordingly, we extract any GenAI-related job posts that contain these model specific job skills or keywords.

Overall, the nine listed GenAI models cover 23 skills (29.48\% of GenAI-related skills) and 7,519 job posts (51.36\% of GenAI-related job posts).
Figure~\ref{fig:genai_post_cnt} presents the distribution of job post counts relevant to the nine models. By far, the most popular is ChatGPT (32.05\% GenAI-related job posts are ChatGPT-oriented). ChatGPT possesses $1.7\times$ job post count as many as the total job post count for all other eight GenAI models. Furthermore, compared to non-GPT-GenAI job posts, GPT-related job posts have \textit{more} bid counts (25.0 vs. 20.0 medians), and higher bid prices (162.97 vs. 115.56 in USD). Thus, our findings confirm the prominence of ChatGPT in the GenAI market on Freelancer.com.

\section{Profiling Employers and bidders relationships}
\label{sec:rq2}

We next seek to understand the relationship between employers and bidders (\textbf{RQ2}).
Out of the 3,893,131 users collected, we concentrate on the 1,365,543 (35.08\%) \textit{active} employers and bidders who have submitted either at least three job posts or at least three bids.

\pb{Inducing bid network.} 
We induce a bid network   
based on users' bidding actions. If a \emph{bidder} \emph{bids} on a \emph{job post} submitted by an \emph{employer}, we assign a directed edge from the bidder to the employer. Each edge is weighted by the total number of bids the bidder has sent for jobs posted by that by employer.
Accordingly, our bid network is a directed graph including 3,032,747 nodes (active users and their neighbors) and 45,949,647 edges. Note, this bid network of active users covers 77.90\% of all users and 97.21\% bidding interactions.

\begin{table}[t]
\centering
\resizebox{.9\linewidth}{!}{%
\begin{tabular}{|L{1em}|L{7.5em}|C{7em}|C{3.5em}|C{9em}|}
\toprule
& \multirow{2}{*}{\textbf{Metrics}} & \multicolumn{2}{c}{\textbf{MWU test}} & \textbf{Mean diff.} \\
\cmidrule(lr){3-4}\cmidrule(lr){5-5}
 &  & $\bm{U}$ & \textbf{p-value} & \textbf{have vs. haven't} \\ \midrule
\multirow{6}{*}{\rotatebox[origin=c]{90}{\textbf{Employers}}} & In-degrees & 582429299.5 & *** & $626.702>174.655$ \\
 & Betweenness & 389076720.0 & *** & $9.716e^{-6}>1.485e^{-6}$ \\
 & PageRank & 543710439.0 & *** & $1.170e^{-5}>3.034e^{-6}$ \\ \cmidrule(l){2-5} 
 & Overall reputation & 321740063.5 & *** & $2.969>2.938$ \\
 & Professionalism & 317332457.5 & *** &  $2.970>2.938$\\
 \midrule
\multirow{7}{*}{\rotatebox[origin=c]{90}{\textbf{Bidders}}} & Out-degrees & 70860555773.0 & *** & $489.623>9.938$\textcolor{white}{0} \\
 & Betweenness & 49138977604.5 & *** & $2.427e^{-6}>7.135e^{-8}$ \\
 & PageRank & 49235232390.5 & *** & $6.233e^{-7}>2.205e^{-7}$ \\ \cmidrule(l){2-5} 
 & Overall reputation & 35627512335.0 & *** & $1.397>0.248$ \\
 & Work quality & 35627080981.0 & *** & $1.397>0.248$ \\
 & Professionalism & 35624818513.0 & *** & $1.399>0.249$ \\
 \bottomrule
\end{tabular}%
}
\caption{Comparison results on metrics profiling users' difference on their social networks and work reputations. ``have vs. haven't'' denotes comparing corresponding metrics for users who have ever submitted or bid any GenAI-related job posts against those who haven't. ***: $\bm{p<0.001}$.}
\label{tab:users}
\end{table}

\pb{Metrics for assessing user difference.}
We rely on the following metrics to measure the difference between users in the bid networks: \textit{in/out-degree}, \textit{betweenness},\footnote{Due to the large scale of the bid network, each node's betweenness is estimated as an average of the betweenness of 100 random sampled subgrapghs involving 0.1\% nodes.} and
\textit{PageRank centrality}.
We are also curious to understand how these patterns relate to the quality of work. Hence, we further inspect the \emph{rating} (0 to 5) of both employers and bidders'
which pertains to their overall reputation, professionalism, and bidders' work quality.

\pb{GenAI is promoted by central and reputable users.}
We first inspect whether users engaged in GenAI-related jobs exhibit distinct centrality locations within the bid network, and distinct reputation scores.
We categorize employers and bidders into two groups based on whether they have ever released/bid on GenAI-related job posts, or not.
We then calculate in-group difference for aforementioned metrics. 
Table~\ref{tab:users} summarizes the inter-group differences for the metrics outlined above reported by the MWU test. 
Our findings indicate that users involved in GenAI-related job posts \textit{do} exhibit distinct network positions and work reputations.
Most notably, users engaged in GenAI-related jobs are a small set of active users (6.03\% of the total active user base) who are reputable and possess higher network centrality.

In more detail, users who have submitted or bid on GenAI-related job posts, on average, show significantly ($p<0.001$) larger in-/out-degree ($3.59\times$ in-degree/$49.27\times$ out-degree of employers/bidders who haven't), higher betweenness (employers: $6.54\times$; bidders: $34.02\times$), and higher PageRank centrality (employers: $3.86\times$; bidders: $2.83\times$) compared to those who have not. This suggests that these users act as brokers and connectors within the bid network. Interestingly, they tend to be employers receiving many bids and active bidders contributing a large number of bids. 

Moreover, users engaged in GPT-related jobs are more reputable, on average.
Users who have submitted or bid on GPT-related job posts on average hold significantly ($p<0.001$) higher overall reputation (employers: $1.01\times$; bidders: $5.63\times$), higher levels of professionalism (employers: $1.01\times$; bidders: $5.63\times$), better work quality (bidders: $5.62\times$).
This indicates that these users are associated with better work reputations, being evaluated as more professional and offering higher-quality services.

\begin{figure}[t]
  \centering
  \includegraphics[width=.9\linewidth]{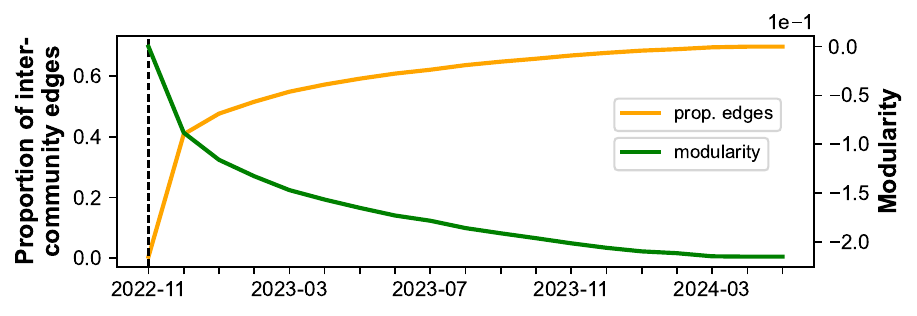}
  \caption{Time series of monthly network modularity and proportion of inter-community edges between users who have ever submitted or bid any GenAI-related job posts and those who haven't. The black dash line denotes ChatGPT's release}
  \label{fig:modularity}
\end{figure}

\pb{Users engaged in GenAI-related jobs are not isolated.}
We conjecture that ``partitions'' may form in the bid graph, whereby some bidders (and jobs) are entirely separated from those who those who have never used GenAI. This divide could make it difficult for individuals without experience in GenAI-related jobs to connect with (or be recommended to) those who have experience in such jobs~\cite{ge2020understanding}.
For this, we inspect the monthly changes of \emph{division strength} between both employers and bidders who have ever submitted or bid on any GenAI-related job posts and those who have not. We quantify such division strength using the network modularity~\cite{newman2004finding} and the proportion of inter-community edges between the two user groups. 

Figure~\ref{fig:modularity} illustrates the monthly time series of modularity and inter-community edge proportion metrics. 
In contrast to our concerns, the community of employers and bidders engaged in GenAI-related jobs is \emph{not} isolated. In fact, we find a \emph{decreasing} trend in modularity, and an increasing proportion of inter-community edges after ChatGPT's release. Surprisingly, this suggests that both employers and bidders, who have submitted or bid on any GenAI-related jobs, consistently establish new connections to those who have not. Accordingly, the strength of division between these two kinds of users is shrinking. Thus, the employers and bidders who have adopted GenAI in freelance jobs are well connected in the bid network, and also improving connectivity with other users without any experience in GenAI-related jobs.

\section{Exploring uses of ChatGPT}
\label{sec:rq3}

Given its leading role on Freelancer.com (\S\ref{sec:rq2}), we next focus on how one particular GenAI model is used in the market: ChatGPT.
Thus, we explore the main job uses of ChatGPT on Freelancer.com, and explore whether employers and bidders' behaviors change after using ChatGPT (\textbf{RQ3}). 

\pb{Job content related to ChatGPT.}
We start by performing a topic analysis using the titles and work descriptions job posts that include the need for workers to use ChatGPT.
We employ BERTopic~\cite{grootendorst2022bertopic} configured with embedding model \texttt{paraphrase-multilingual-mpnet- base-v2}~\cite{reimers-2019-sentence-bert} and a representation model KeyBERT~\cite{grootendorst2020keybert}.
We use this to group job posts into topical clusters. The minimum topic size is set as 15 to ensure coherent clusters. In total, BERTopic analyzes 4,650 GPT-related job posts and groups them into 44 topics, with 1,728 job posts grouped into an outlier topic. We focus on the top 20 most frequent topics (detailed in Figure~\ref{fig:topic_bars}), covering 78.54\% job posts from non-outlier topics. 


Figure~\ref{fig:topic_distance} presents the inter-topic distance map and Figure~\ref{fig:topic_vis} visualizes the emdeddings and the 20 topical clusters of GPT-related job posts.
We observe an obvious division between two groups of topical clusters, indicating the existence of 2 main underlying uses of ChatGPT among employers. Through a qualitative analysis of the 20 topics and their corresponding representative job posts, we characterize the employers' requirements for using ChatGPT in 2 categories:

\begin{itemize} [leftmargin=*]
    \item \textbf{Use 1: Integrating ChatGPT into software development.} 
    A common goal of employers is to ask for workers who can integrate ChatGPT into specific software and web applications. This target pertains to 13 topics, covering 1,560 (62.02\%) GPT-related job posts from non-outlier topics.
    
    \item \textbf{Use 2: Reproducing human-generated content.} 
    Another common goal of employers is to recruit workers who can use ChatGPT to reproduce human-generated content. This theme contains 7 topics, covering 735 (25.15\%) analyzed job posts. 
\end{itemize}
Appendix~\ref{appd:job_content} also provides a more detailing instruction with topic-wise statistics. 

\begin{figure*}[t]
    \centering
    \begin{subfigure}[b]{.24\linewidth}
        \centering
        \includegraphics[width=\linewidth]{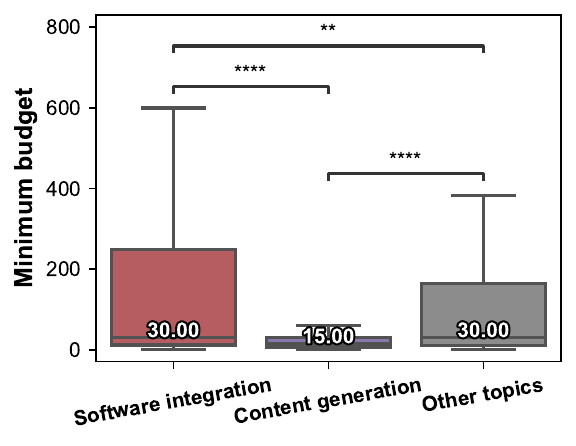} 
    \end{subfigure}
    \begin{subfigure}[b]{.24\linewidth}
        \centering
        \includegraphics[width=\linewidth]{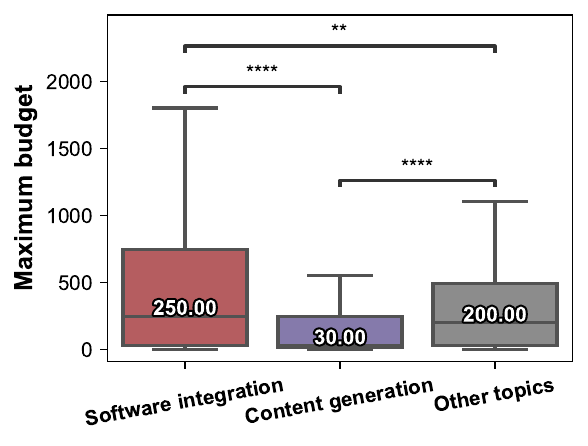}  
    \end{subfigure}
    \begin{subfigure}[b]{.24\linewidth}
        \centering
        \includegraphics[width=\linewidth]{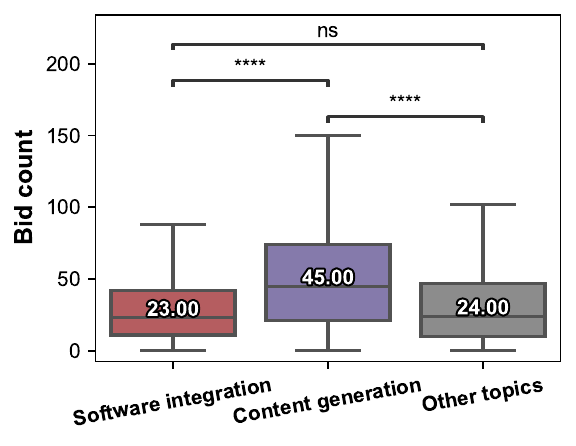}  
    \end{subfigure}
    \begin{subfigure}[b]{.24\linewidth}
        \centering
        \includegraphics[width=\linewidth]{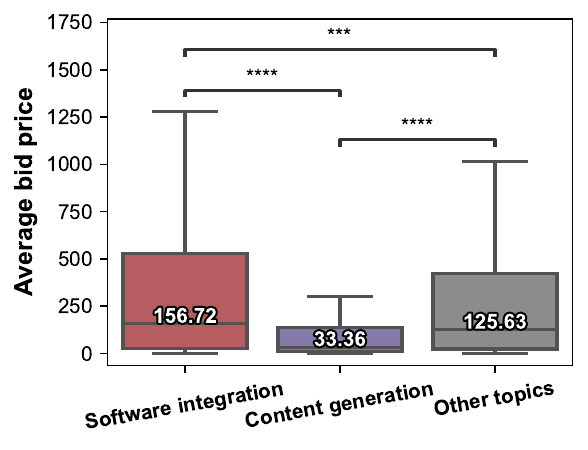} 
    \end{subfigure}
    \caption{Comparison of minimum/maximum budget, bid count, and average bid price between GPT-related job posts within diverse targets. The statistical significance is reported by the Kruskal-Wallis test with Dunn's post-hoc test; ****: $\bm{p<0.0001}$, ***: $\bm{p<0.001}$ **: $\bm{p<0.01}$, *: $\bm{p<0.05}$.}
    \label{fig:target_compare}
\end{figure*}

\pb{Diverse uses of ChatGPT trigger different behaviors.}
We next explore whether users focusing on the above topics demonstrate distinct job posting and bidding patterns. For this, we compare the minimum/maximum budgets, bids' count and price of job posts from different topical groups. Figure~\ref{fig:target_compare}
illustrates the comparison results for job posts using ChatGPT for software integration, content generation, and other topics. The results reveal that users do present significantly different investment and bidding patterns when they use ChatGPT for different goals.

We see that employers invest more budgets for software integration compared to other uses of ChatGPT. Job posts using ChatGPT for software integration, on average, possess significantly more budget than either content generation ($3.82\times$/$2.87\times$ mean values of minimum/maximum budget of content generation job posts, $p<0.001$) or other topics (minimum/maximum budget: $1.57\times$/$1.11\times$, $p<0.01$).
Unsurprisingly, this attracts bidders who demand higher payments (i.e., a higher bid price) when biding for software integration jobs, as compared to content generation jobs ($3.85\times$, $p<0.001$) or others ($1.42\times$, $p<0.001$). 

That said, there are more bidders who bid for jobs using ChatGPT to reproduce human-generated content (median 45 bids per job post vs. 23 for software integration jobs), as compared to software development. 
Although associated with the lowest level of budgets, job posts that ask that ChatGPT is used for content generation receive a significantly ($p<0.001$) larger volume of bids than either software integration ($1.74\times$) or other topics ($1.61\times$). We assume this is because content generation is easier for a wide array of bidders (compared to software integration).
Alternatively, given that a positive Pearson correlation is found between the required job skills count and bids count ($r=0.115, p<0.001$) in job posts, this could be because jobs using ChatGPT for content generation cover a broader range of job skills. 
This means they could attract more bidders from different industries. Indeed, we see the content generation jobs require significantly ($p<0.001$) more job skills from more diverse industries.

\pb{Using ChatGPT catalyzes more job posting.}
One potential reason for the growing popularity of ChatGPT is that employers are eager to post more GPT-related jobs after previously using ChatGPT, and seeing its benefits.
Thus, we next explore if this effect exists in the freelancing market.
We focus on 603 active employers who have released at least 3 job posts both before and after their first posting GPT-related job. According to the Wilcoxon signed-rank test, we find a significant ($p<0.001$) increase in employers' average daily job posts following their first GPT-related job, with the median rising from 0.045 to 0.065. 
This suggests that utilizing ChatGPT drives employers to post jobs more frequently.
Moreover, we find that on average 21.25\% of job posts after each employer's first GPT-related job continue to require ChatGPT. In fact, 7.63\% of these employers demonstrate significant reliance on ChatGPT, with more than 50\% of their job posts requiring its use. We therefore posit that ChatGPT is fostering greater reliance among employers, catalyzing them to release more jobs that use ChatGPT. 

\pb{GPT-related jobs encourage expanding job skills.} 
It has been argued that the use of GenAI may impact users' skills and training~~\cite{10.1145/3613904.3642700, 10.1145/3706598.3714035}.
For example, the above suggests that it will become increasingly necessary for users to improve their ChatGPT related skills to remain competitive (\S\ref{sec:RQ1}). 
Motivated by this, we next explore whether bidders' job skill sets significantly change after they bidding GPT-related jobs.
We focus on 22,766 active bidders who have submitting at least 3 bids both before and after their first bid on a GPT-related job. 
The Wilcoxon signed-rank test reports that bidders' distinct job skill count significantly ($p<0.001$) increases after their first time biding on a GPT-related job, with the median rising from 59.0 to 66.0. Additionally, these job skills seem to come from slightly more diverse industries, with the median distinct industry count staying as 8.0 but the mean rising from 8.01 to 8.49 ($p<0.001$). This indicates that GPT-related jobs may play a role in motivating bidders to expand their job skills and engage in diverse jobs from different industries, reflecting its potential role in helping to remove the skill barriers between bidders and unfamiliar job domains.

\section{Profiling Skill Demands Relevant to ChatGPT}
\label{sec:rq4}

There has long been interest among jobs seekers, regulators and investors in identifying which job skills will remain relevant or become obsolete in the future. This has become particular relevant since the market disruptions driven by GenAI over the last two years~\cite{jain2023impact, giordano2024impact}. 
Indeed, our prior results show that certain job skills (e.g., ``Machine Learning'') have become more in-demand with in GenAI jobs.
With this in-mind, we take ChatGPT (the most popular GenAI technology on Freelancer.com) as a case study and explore the predictability of what job skills will become most needed (or not) for GPT-related jobs (\textbf{RQ4}). 

\begin{table}[]
\centering
\resizebox{.9\linewidth}{!}{%
\begin{tabular}{|L{2.5em}|L{8em}|C{4em}|C{3.5em}|C{10em}|}
\toprule
& \multirow{2}{*}{\textbf{Metrics}} & \multicolumn{2}{c}{\textbf{MWU test}} & \textbf{Mean diff.} \\
\cmidrule(lr){3-4}\cmidrule(lr){5-5}
 &  & $\bm{U}$ & \textbf{p-value} & \textbf{relevant vs. irrelevant} \\ \midrule
\multirow{5}{*}{\rot{\textbf{\Tr{Overall\\distribution}}}} & Job post count & 602804.0 & *** & $6789.60 > 396.93$\textcolor{white}{0} \\
& Minimum budget & 308523.5 & 0.005 & $30.18 < 32.79$ \\
& Maximum budget & 313826.5 & 0.049 & $100.43 < 118.04$ \\
& Bid count per post & 426615.0 & *** & $9.52 > 6.26$ \\
& Bid price & 333387.0 & 0.965 & $85.78 < 88.73$ \\ \midrule
\multirow{5}{*}{\rot{\textbf{\Tr{Median level\\(daily avg.)}}}} & Job post count & 585052.5 & *** & $20.19 > 1.66$\textcolor{white}{0} \\
& Minimum budget & 522220.5 & *** & $94.92 > 46.79$ \\
& Maximum budget & 467794.0 & *** & $228.66 > 152.89$ \\
& Bid count per post & 516312.0 & *** & $15.76 > 8.24$\textcolor{white}{0} \\
& Bid price & 487975.0 & *** & $223.88 > 112.09$ \\ \midrule
\multirow{5}{*}{\rot{\textbf{\Tr{Trends\\(daily avg.)}}}} & Job post count & 215474.0 & *** & $-0.021 < 0.0002$\textcolor{white}{-} \\
& Minimum budget & 326510.0 & 0.473 & $-0.498 < -0.188$ \\
& Maximum budget & 336181.0 & 0.819 & $-6.03e^{7} < -1.59e^{7}$ \\
& Bid count per post & 456054.0 & *** & $0.019 > 0.006$ \\
& Bid price & 314787.0 & 0.062 & $-4.75e^{6} < -2.65e^{6}$ \\ \bottomrule
\end{tabular}%
}
\caption{Comparison results on features profiling job skills' difference before ChatGPT's release. ``Mean diff.'' denotes the comparison results of the mean value of corresponding metrics between two groups; ***: $\bm{p<0.001}$.}
\label{tab:time_series_compare}
\end{table}

\pb{Feature engineering.} 
We first try to curate a feature set that can effectively characterize the differences between the job skills that will remain relevant in the ChatGPT era and those that are likely to be phased out. 
For the 1,871 existing job skills, a skill is classified as \textbf{relevant} to the ChatGPT market if it is required by at least 3 GPT-related job posts; otherwise, it is classified as \textbf{irrelevant}.
We experiment with key features inspired by recent market forecast research~\cite{patel2024systematic, dawson2020predicting,shi2020salience}. For each skill, we compute the following features: its total job post count, the median values of minimum/maximum budget, bid count per post, and bid price, the median values and trends of daily job post count, daily average minimum/maximum budget, bid count per post, and bid price.
Note, all features are measured based on the market data before ChatGPT's release and those related to the trends of time series are estimated as the regression coefficient as \(Daily~observations=Trend*Days+Intercept+\epsilon\).

Table~\ref{tab:time_series_compare} summarizes the comparison results using the Mann-Whitney U (MWU) test for these features. Among the fifteen features, nine of them demonstrate significant differences ($p<0.001$) between the two job groups. This suggests that these nine features may effectively differentiate in-demand job skills for GPT-related jobs. 

\pb{Profiling skills in-demand in ChatGPT jobs.} 
To understand their traits, we briefly characterize the above features (Table~\ref{tab:time_series_compare}).
We see that many skills relevant to the ChatGPT job market were already popular before ChatGPT's release.
These skills possess significantly ($p<0.001$) more job posts ($17.11\times$ average and $11.90\times$ median value of the job post count of skills exiting from ChatGPT market), alongside higher median bid counts per job post ($\mu$: $1.52\times$, $mid$: $1.60\times$). Additionally,
before ChatGPT's release, they hold significantly ($p<0.001$) higher median daily observations of job post count ($\mu$: $12.17\times$, $mid$: $5.00\times$), average minimum budget ($\mu$: $2.03\times$, $mid$: $2.58\times$), average maximum budget ($\mu$: $1.50\times$, $mid$: $1.38\times$), average bid count per job post ($\mu$: $1.91\times$, $mid$: $1.94\times$), and average bid price ($\mu$: $1.50\times$, $mid$: $1.38\times$). 
These results imply that the emerging ChatGPT market's demand towards job skills is selective. GPT-related jobs seem more likely to derive from highly-demanded and well-paid professions in freelancing market. 
These professions mainly represent skills from industries like ``IT, Software'' (e.g., ``PHP'', ``HTML'', and ``WordPress''), ``Design, Media'' (e.g., ``Graphic Design'', ``Website Design'', and ``Photoshop''), ``Writing'' (e.g., ``Article Writing'', ``Content Writing'', ``Research Writing''), ``Engineering'' (e.g., ``Engineering'' and ``Machine Learning''), and ``Sales'' (e.g., ``Internet Marketing'', ``Social Media Marketing'', and ``Advertising''), which closely align with the content of employers' uses of ChatGPT (showed in \S\ref{sec:rq3}).


\begin{figure*}[t]
    \centering
    \begin{subfigure}[b]{.24\linewidth}
        \centering
        \includegraphics[width=\linewidth]{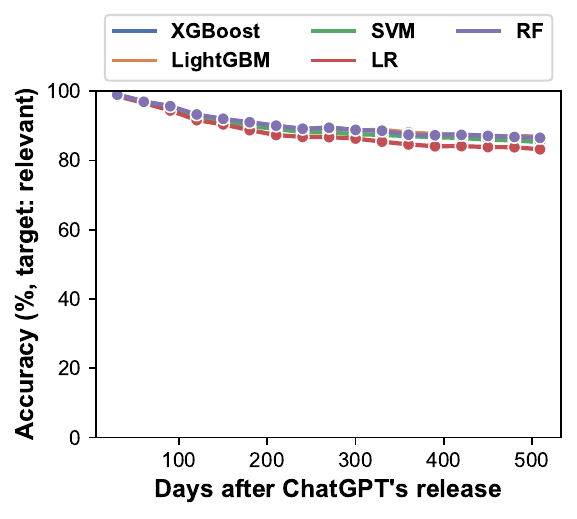}  
    \end{subfigure}
    \begin{subfigure}[b]{.24\linewidth}
        \centering
        \includegraphics[width=\linewidth]{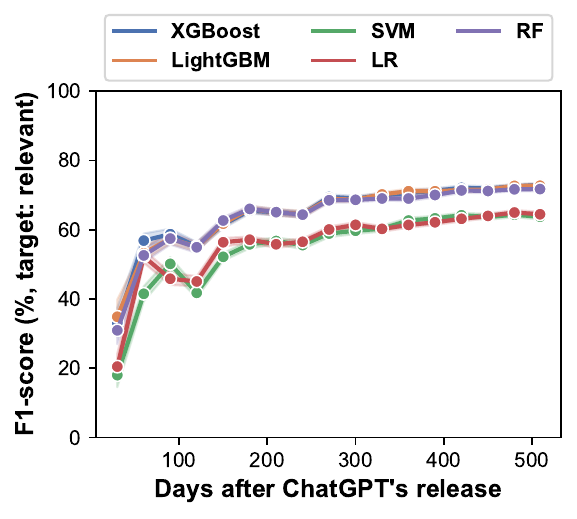}  
    \end{subfigure}
    \begin{subfigure}[b]{.24\linewidth}
        \centering
        \includegraphics[width=\linewidth]{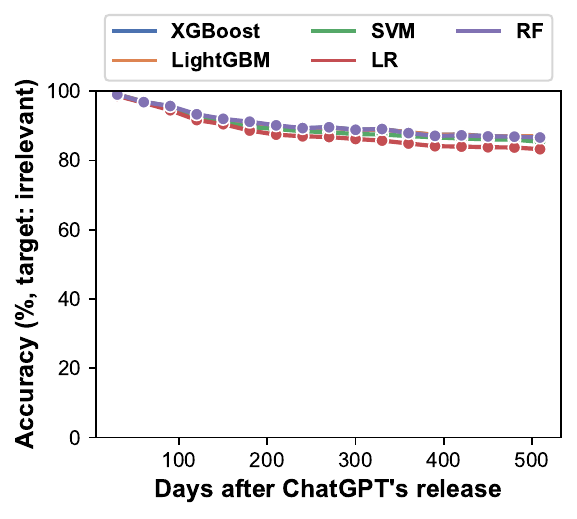}  
    \end{subfigure}
    \begin{subfigure}[b]{.24\linewidth}
        \centering
        \includegraphics[width=\linewidth]{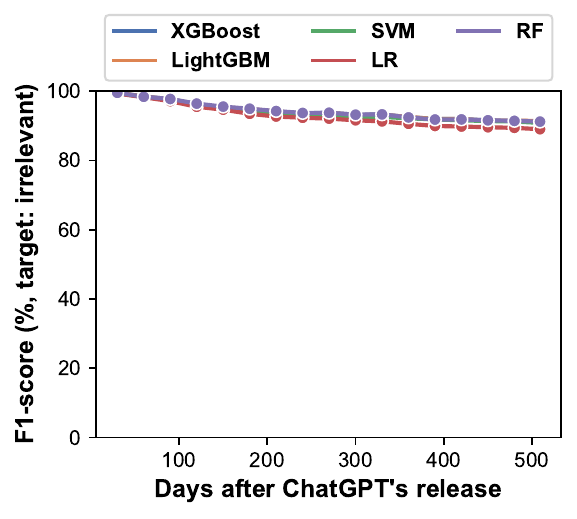}  
    \end{subfigure}
    \caption{Classifiers performance in defining whether job skills are relevant or irrelevant to ChatGPT market for every 30 days after ChatGPT's release.}
    \label{fig:early_market}
\end{figure*}

\begin{table}[]
\centering
\resizebox{.9\linewidth}{!}{%
\begin{tabular}{|l|cc|cc|}
\toprule
\multirow{2}{*}{\textbf{Classifier}} & \multicolumn{2}{c|}{\textbf{Target: relevant}} & \multicolumn{2}{c|}{\textbf{Target: irrelevant}} \\ 
\cmidrule(lr){2-3}\cmidrule(lr){4-5}
& Accuracy & F1-score & Accuracy & F1-score \\\midrule
{SVM} & $0.853_{\pm0.0145}$ & $0.637_{\pm0.0436}$ & $0.854_{\pm0.0118}$ & $0.908_{\pm0.0070}$ \\ \midrule
{RF} & $0.865_{\pm0.0146}$ & $0.717_{\pm0.0321}$ & $0.866_{\pm0.0162}$ & $0.912_{\pm0.0108}$ \\ \midrule
{LR} & $0.832_{\pm0.0160}$ & $0.645_{\pm0.0365}$ & $0.832_{\pm0.0129}$ & $0.890_{\pm0.0088}$\\\midrule
{XGBoost} & \bm{$0.868_{\pm0.0145}$} & \bm{$0.729_{\pm0.0301}$} & \bm{$0.868_{\pm0.0153}$} & \bm{$0.913_{\pm0.0102}$} \\ \midrule
{LightGBM} & $0.867_{\pm0.0147}$ & $0.727_{\pm0.0318}$ & {$0.867_{\pm0.0147}$} & {$0.912_{\pm0.0097}$} \\ \bottomrule
\end{tabular}%
}
\caption{Classifiers' performance in classifying job skills relevant or irrelevant to GPT-related jobs.}
\label{tab:prediction}
\end{table}

\pb{Identifying in-demand skills by curated feature set.}
We finally test how predictable the set of ChatGPT-relevant skills are.
For this, we create an input feature set containing the nine significant features (above), and experiment with 
five common machine learning algorithms: support vector machine (SVM), random forest (RF), logistic regression (LR), XGBoost~\cite{chen2016xgboost}, and LightGBM~\cite{ke2017lightgbm}. 
The task is to classify which job skills will remain relevant/irrelevant for future ChatGPT jobs based on their pre-GPT-release statistics. We employ the default parameter setting of all experimented model and present them in Table~\ref{tab:models}.

\pb{Classifiers' performance.} 
{Table~\ref{tab:prediction} summarizes the performance of different classifiers. The best classifier is XGBoost, on average achieving 0.868 accuracy with 0.729 F1-score and 0.868 accuracy with 0.913 F1-score for classifying the relevant and irrelevant respectively. Such results validates the effectiveness of our characterization, suggesting that our selected features provide meaningful signals for predicting skill relevance in the evolving GenAI job market. The strong performance of XGBoost, particularly in identifying irrelevant skills (F1-score of 0.913), indicates that our model can reliably distinguish declining skills, offering actionable insights for both job seekers and employers.}

\pb{The stable periods of job demands for early market.}
{One potential application of above classification is to help define when the skill demand of ChatGPT market become stable, providing an empirical understanding of the maturation timeline of the emerging technology job markets. To explore this, we use the five trained classifier to predict relevant and irrelevant skills for every 30-day interval after ChatGPT's release. We consider that when the prediction performance starts converging, the skill demand of ChatGPT market turns into stable.}

Figure~\ref{fig:early_market} shows the distribution of classifiers' accuracy and F1-score in predicting the relevant and irrelevant skills under diverse periods. We find, for all experimented classifiers, the F1-score always quickly converges to the value of the end day of our data collection. 
This suggests the prediction based on our characterization can become coherent at a very early stage of the ChatGPT market. Consequently, job seekers and investors can utilize such results for decision making. Based on this, we next assess the earliest period (in a 30-day interval) when the F1-score converges into a range within a $3\%$ deviation of the value of end day of our data collection, offering a lens to when the prediction existing job skills in ChatGPT market has become stable and accurate.
Additionally,  illustrates 
We also summarize the earliest periods when the classifiers achieve a stable prediction for relevant/irrelevant skills in Table~\ref{tab:early_market}. 
Our results suggest that even for rapidly evolving GenAI technologies (e.g., ChatGPT), market forces follow observable patterns that can inform timely, data-driven decisions.

\section{Related Work}

\pb{GenAI and labor market.} There has been growing interest in the relationship between GenAI and the existing labor market. The employment of GenAI in jobs has already been studies in various industries, including education~\cite{pang2024artificialhumanlecturersinitial, 10.1145/3573051.3596191}, art~\cite{wei2024understandingimpactaigenerated, 10.1145/3635636.3664263}, and sales~\cite{chui2023economic}. These indicate a trend that new jobs are combining both human and ChatGPT intelligence. 
Our study confirms this observation within the freelancing market.
Recently, researchers have argue the needs for measuring the evolution of online labor market relevant to GenAI,
in order to understand GenAI's role on digital economy~\cite{10.1145/3589335.3641295, KSHETRI2024102716}. 
Nowadays, freelancing platforms serve as a vital source of employment. {In this paper, we perform a platform-wise study of GenAI's role on freelancing markets}. In contrast to prior research~\cite{dolata2024development, woodruff2024knowledge}, {our work provides a targeted understanding of the role of GenAI in leading the rise of GenAI-related jobs on the freelancing market}, profiling the job posting patterns and users engaging.

\pb{ChatGPT's impact on freelancing platforms.}
A small set of recent studies have inspected the impact of ChatGPT on freelancing platforms. Liu et al. find that introducing ChatGPT leads to a decline in labor demand for text related and programming-related jobs~\cite{liu2023generate}, while Demirci et al. show that jobs requiring manual-intensive skills are less likely to be replaced by ChatGPT~\cite{demirci2023ai}.
In contrast to these work, which focus on certain job types, our study provides a broader view to the job content related to ChatGPT. Moreover, we further inspect the effect of ChatGPT usages in shifting users' behaviors on the freelancing platform.
Other researchers try to profile ChatGPT's impact on job skill transitions~\cite{liu2023generate, yiu2024ai}, and labor substitution~\cite{yilmaz2023ai, demirci2023ai}. They also explore the potential benefits or risks~\cite{george2023chatgpt, meyer2023chatgpt} for freelancing platforms. Our research complements prior studies by offering a more comprehensive empirical analysis on a large-scale dataset. More importantly, we provide a novel angle to profile and predict skills relevant to the ChatGPT market, providing a novel lens for investors and job skill seekers to understand the evolution of the ChatGPT-driven labor market.


\section{Conclusion \& Discussion}

{Our study provides a large-scale empirical analysis of the role of GenAI on the online freelancing market, with a particular focus on the period following the release of ChatGPT. Our findings reveal several important trends and dynamics that have implications to facilitate research communities to better understand the broader gig economy and design platform policies in the GenAI era.}

\pb{The dual nature of GenAI's role.}
{Contrary to popular narratives that frame GenAI primarily as a disruptive or job-replacing force~\cite{10.1145/3613904.3642700, 10.1145/3706598.3714035}, our results suggest a more nuanced reality. While we do observe a decline in non-GenAI job posts over time, the emergence of GenAI-related jobs has not only compensated for this decline but has also introduced new forms of labor demand. Jobs requiring GenAI integration --- particularly in software integration --- are not only more numerous but also better compensated.
This indicates that GenAI is acting as a catalyst for skill transformation rather than a simple substitute for human labor. Freelancers who adapt to these technologies are likely to benefit from higher pay, greater demand, and opportunities to expand their skill sets across industries. Notably, ChatGPT has emerged as the most popular GenAI technology within the freelancing ecosystem, accounting for nearly one-third of all GenAI-related job posts. Its release served as a significant inflection point, accelerating demand for GenAI skills and reshaping employer and bidder behaviors. The fact that employers who use ChatGPT once are likely to continue using it --- and even increase their job posting frequency --- suggests that ChatGPT is becoming embedded in freelance workflows. This has implications for both platform designers and policymakers: supporting upskilling and integration tools around leading AI models may help mitigate skill gaps and ensure equitable access to new opportunities.}

\pb{The evolution of GenAI-related freelancer community.}
{Our analysis reveals that the community of freelancers engaged with GenAI, particularly following the release of ChatGPT, is not a isolated niche. We observe that GenAI users occupy central positions within the bid network, acting as hubs that connect diverse groups of employers and bidders. Their high betweenness and PageRank centrality scores suggest they play a critical role in facilitating information and opportunity flow across the platform.
This finding counters concerns that AI-adopting users might form exclusionary clusters, potentially marginalizing those without AI experience. }

{Notably, the structural integration of GenAI users has increased over time. Following ChatGPT’s release, network modularity decreased while the proportion of edges between GenAI and non-GenAI communities grew. This indicates that GenAI-related projects are increasingly served by a mixed workforce, where users with and without AI experience collaborate rather than compete in isolation. 
This integration may be driven by the multidisciplinary nature of GenAI projects.
For example, a software development job requiring AI integration may also need UI designers, copywriters, or project managers who do not directly work with AI.}

{These dynamics highlight an important shift that the GenAI freelancer community has become an integrative force within the digital labor market.
Platforms can reinforce this positive evolution by promoting collaborative project structures, offering blended skill certifications, and ensuring that AI-related opportunities are visible and accessible to all users, regardless of their prior AI experience.}

\pb{Limitation \& Future work.}
There are several areas where our study could be enhanced. Currently, our research focuses exclusively on Freelancer.com. In the future, we aim to include other freelancing platforms, such as Upwork and Fiverr, to gain a broader perspective on the role of GenAI. Additionally, part of our analysis has centered on ChatGPT, given its leading position in the GenAI landscape. Moving forward, we plan to gather more market data related to other GenAI models (i.e., Llama, Gemini) and extend our analysis to encompass these alternatives.

%
\bibliographystyle{ACM-Reference-Format}
\bibliography{sample-base}

@inproceedings{10.1145/3706598.3714035,
author = {Varanasi, Rama Adithya and Wiesenfeld, Batia Mishan and Nov, Oded},
title = {AI Rivalry as a Craft: How Resisting and Embracing Generative AI Are Reshaping the Writing Profession},
year = {2025},
isbn = {9798400713941},
publisher = {Association for Computing Machinery},
address = {New York, NY, USA},
url = {https://doi.org/10.1145/3706598.3714035},
doi = {10.1145/3706598.3714035},
booktitle = {Proceedings of the 2025 CHI Conference on Human Factors in Computing Systems},
articleno = {1198},
numpages = {19},
location = {
},
series = {CHI '25}
}

@inproceedings{10.1145/3613904.3642700,
author = {Woodruff, Allison and Shelby, Renee and Kelley, Patrick Gage and Rousso-Schindler, Steven and Smith-Loud, Jamila and Wilcox, Lauren},
title = {How Knowledge Workers Think Generative AI Will (Not) Transform Their Industries},
year = {2024},
isbn = {9798400703300},
publisher = {Association for Computing Machinery},
address = {New York, NY, USA},
url = {https://doi.org/10.1145/3613904.3642700},
doi = {10.1145/3613904.3642700},
booktitle = {Proceedings of the 2024 CHI Conference on Human Factors in Computing Systems},
articleno = {641},
numpages = {26},
location = {Honolulu, HI, USA},
series = {CHI '24}
}

@article{newman2004finding,
  title={Finding and evaluating community structure in networks},
  author={Newman, Mark EJ and Girvan, Michelle},
  journal={Physical review E},
  volume={69},
  number={2},
  pages={026113},
  year={2004},
  publisher={APS}
}

@misc{freelancer-report,
  author       = {Freelancer.com},
  title        = {2022 Annual Report},
  year         = {2023},
  url          = {https://www.freelancer.com/about/investor-pdf.php?id=194473978&name=FLN+ANNUAL_REPORT_2022&w=f&redirect-times=1&ngsw-bypass=}
}

@misc{freelancer-report2,
  author       = {Skillademia.com},
  title        = {Freelancer.com Statistics (2024): User Growth, Revenue, Demographics, Top Skills in Demand, and AI},
  year         = {2024},
  url          = {https://www.skillademia.com/statistics/freelancer-com-statistics/#:~:text=Freelancer.com%20Job%20Postings%20Data,the%20platform%20received%2041%20bids}
}

@misc{gig-is-up,
  author       = {Harvard Business School},
  title        = {The Gig Is Up},
  year         = {2020},
  url          = {https://news.harvard.edu/gazette/story/2020/11/during-covid-19-remote-freelance-work-is-on-the-rise/}
}

@misc{gpt-kick-genai,
  author       = {CNBC},
  title        = {AI agents are having a 'ChatGPT moment' as investors look for what’s next after chatbots},
  year         = {2024},
  url          = {https://www.cnbc.com/2024/06/07/after-chatgpt-and-the-rise-of-chatbots-investors-pour-into-ai-agents.html}
}

@misc{gen-ai-disrupt-creative,
  author       = {Harvard Business Review},
  title        = {How Generative AI Could Disrupt Creative Work},
  year         = {2023},
  url          = {https://hbr.org/2023/04/how-generative-ai-could-disrupt-creative-work}
}

@inproceedings{shi2020salience,
  title={Salience and market-aware skill extraction for job targeting},
  author={Shi, Baoxu and Yang, Jaewon and Guo, Feng and He, Qi},
  booktitle={Proceedings of the 26th ACM SIGKDD International Conference on Knowledge Discovery \& Data Mining},
  pages={2871--2879},
  year={2020}
}

@inproceedings{dawson2020predicting,
  title={Predicting skill shortages in labor markets: A machine learning approach},
  author={Dawson, Nikolas and Rizoiu, Marian-Andrei and Johnston, Benjamin and Williams, Mary-Anne},
  booktitle={2020 IEEE International Conference on Big Data (Big Data)},
  pages={3052--3061},
  year={2020},
  organization={IEEE}
}

@article{patel2024systematic,
  title={A Systematic Review on Graph Neural Network-based Methods for Stock Market Forecasting},
  author={Patel, Manali and Jariwala, Krupa and CHATTOPADHYAY, CHIRANJOY},
  journal={ACM Computing Surveys},
  year={2024},
  publisher={ACM New York, NY}
}

@article{giordano2024impact,
  title={The impact of ChatGPT on human skills: A quantitative study on twitter data},
  author={Giordano, Vito and Spada, Irene and Chiarello, Filippo and Fantoni, Gualtiero},
  journal={Technological Forecasting and Social Change},
  volume={203},
  pages={123389},
  year={2024},
  publisher={Elsevier}
}

@article{jain2023impact,
  title={The Impact Of Chatgpt On Job Roles And Employment Dynamics},
  author={Jain, Shalu and Khare, Ankur and Goel, OGPP and Singh, SP},
  journal={JETIR},
  volume={10},
  number={7},
  pages={370},
  year={2023}
}

@article{lazaroiu2023generative,
  title={How generative artificial intelligence technologies shape partial job displacement and labor productivity growth},
  author={Lazaroiu, George and Rogalska, El{\.z}bieta},
  journal={Oeconomia Copernicana},
  volume={14},
  number={3},
  pages={703--706},
  year={2023},
  publisher={Instytut Bada{\'n} Gospodarczych}
}

@inproceedings{munoz2022new,
  title={New futures of work or continued marginalization? The rise of online freelance work and digital platforms},
  author={Munoz, Isabel and Sawyer, Steve and Dunn, Michael},
  booktitle={Proceedings of the 1st Annual Meeting of the Symposium on Human-Computer Interaction for Work},
  pages={1--7},
  year={2022}
}

@article{wei2024exploring,
  title={Exploring the Use of Abusive Generative AI Models on Civitai},
  author={Wei, Yiluo and Zhu, Yiming and Hui, Pan and Tyson, Gareth},
  journal={arXiv preprint arXiv:2407.12876},
  year={2024}
}

@article{chui2023economic,
  title={The economic potential of generative AI},
  author={Chui, Michael and Hazan, Eric and Roberts, Roger and Singla, Alex and Smaje, Kate},
  year={2023},
  publisher={McKinsey \& Company}
}

@article{KSHETRI2024102716,
title = {Generative artificial intelligence in marketing: Applications, opportunities, challenges, and research agenda},
journal = {International Journal of Information Management},
volume = {75},
pages = {102716},
year = {2024},
issn = {0268-4012},
doi = {https://doi.org/10.1016/j.ijinfomgt.2023.102716},
url = {https://www.sciencedirect.com/science/article/pii/S026840122300097X},
author = {Nir Kshetri and Yogesh K. Dwivedi and Thomas H. Davenport and Niki Panteli}
}

@inproceedings{10.1145/3589335.3641295,
author = {He, Fengxiang and Du, Mengnan and Filos-Ratsikas, Aris and Cheng, Lu and Song, Qingquan and Lin, Min and Vines, John},
title = {AI Driven Online Advertising: Market Design, Generative AI, and Ethics},
year = {2024},
isbn = {9798400701726},
publisher = {Association for Computing Machinery},
address = {New York, NY, USA},
url = {https://doi.org/10.1145/3589335.3641295},
doi = {10.1145/3589335.3641295},
booktitle = {Companion Proceedings of the ACM Web Conference 2024},
pages = {1407–1409},
numpages = {3},
location = {Singapore, Singapore},
series = {WWW '24}
}

@article{liu2023generate,
  title={"Generate" the Future of Work through AI: Empirical Evidence from Online Labor Markets},
  author={Liu, Jin and Xu, Xingchen and Li, Yongjun and Tan, Yong},
  journal={arXiv preprint arXiv:2308.05201},
  year={2023}
}

@article{yilmaz2023ai,
  title={AI-driven labor substitution: Evidence from google translate and ChatGPT},
  author={Yilmaz, Erdem Dogukan and Naumovska, Ivana and Aggarwal, Vikas A},
  year={2023},
  publisher={INSEAD Working Paper}
}

@article{george2023chatgpt,
  title={ChatGPT and the future of work: a comprehensive analysis of AI'S impact on jobs and employment},
  author={George, A Shaji and George, AS Hovan and Martin, AS Gabrio},
  journal={Partners Universal International Innovation Journal},
  volume={1},
  number={3},
  pages={154--186},
  year={2023}
}

@inproceedings{woodruff2024knowledge,
  title={How knowledge workers think generative ai will (not) transform their industries},
  author={Woodruff, Allison and Shelby, Renee and Kelley, Patrick Gage and Rousso-Schindler, Steven and Smith-Loud, Jamila and Wilcox, Lauren},
  booktitle={Proceedings of the CHI Conference on Human Factors in Computing Systems},
  pages={1--26},
  year={2024}
}

@inproceedings{dolata2024development,
  title={Development in times of hype: How freelancers explore Generative AI?},
  author={Dolata, Mateusz and Lange, Norbert and Schwabe, Gerhard},
  booktitle={Proceedings of the IEEE/ACM 46th International Conference on Software Engineering},
  pages={1--13},
  year={2024}
}

@article{meyer2023chatgpt,
  title={ChatGPT and large language models in academia: opportunities and challenges},
  author={Meyer, Jesse G and Urbanowicz, Ryan J and Martin, Patrick CN and O’Connor, Karen and Li, Ruowang and Peng, Pei-Chen and Bright, Tiffani J and Tatonetti, Nicholas and Won, Kyoung Jae and Gonzalez-Hernandez, Graciela and others},
  journal={BioData Mining},
  volume={16},
  number={1},
  pages={20},
  year={2023},
  publisher={Springer}
}

@article{demirci2023ai,
  title={Who is AI Replacing? The Impact of ChatGPT on Online Freelancing Platforms},
  author={Demirci, Ozge and Hannane, Jonas and Zhu, Xinrong},
  journal={The Impact of ChatGPT on Online Freelancing Platforms (October 15, 2023)},
  year={2023}
}

@inproceedings{10.1145/3573051.3596191,
author = {Smolansky, Adele and Cram, Andrew and Raduescu, Corina and Zeivots, Sandris and Huber, Elaine and Kizilcec, Rene F.},
title = {Educator and Student Perspectives on the Impact of Generative AI on Assessments in Higher Education},
year = {2023},
isbn = {9798400700255},
publisher = {Association for Computing Machinery},
address = {New York, NY, USA},
url = {https://doi.org/10.1145/3573051.3596191},
doi = {10.1145/3573051.3596191},
booktitle = {Proceedings of the Tenth ACM Conference on Learning @ Scale},
pages = {378–382},
numpages = {5},
location = {Copenhagen, Denmark},
series = {L@S '23}
}

@misc{pang2024artificialhumanlecturersinitial,
      title={Artificial Human Lecturers: Initial Findings From Asia's First AI Lecturers in Class to Promote Innovation in Education}, 
      author={Ching Christie Pang and Yawei Zhao and Zhizhuo Yin and Jia Sun and Reza Hadi Mogavi and Pan Hui},
      year={2024},
      eprint={2410.03525},
      archivePrefix={arXiv},
      primaryClass={cs.HC},
      url={https://arxiv.org/abs/2410.03525}, 
}

@inproceedings{10.1145/3635636.3664263,
author = {Kawakami, Reishiro and Venkatagiri, Sukrit},
title = {The Impact of Generative AI on Artists},
year = {2024},
isbn = {9798400704857},
publisher = {Association for Computing Machinery},
address = {New York, NY, USA},
url = {https://doi.org/10.1145/3635636.3664263},
doi = {10.1145/3635636.3664263},
booktitle = {Proceedings of the 16th Conference on Creativity \& Cognition},
pages = {79–82},
numpages = {4},
location = {Chicago, IL, USA},
series = {C\&C '24}
}

@misc{wei2024understandingimpactaigenerated,
      title={Understanding the Impact of AI Generated Content on Social Media: The Pixiv Case}, 
      author={Yiluo Wei and Gareth Tyson},
      year={2024},
      eprint={2402.18463},
      archivePrefix={arXiv},
      primaryClass={cs.CY},
      url={https://arxiv.org/abs/2402.18463}, 
}

@article{grootendorst2022bertopic,
  title={BERTopic: Neural topic modeling with a class-based TF-IDF procedure},
  author={Grootendorst, Maarten},
  journal={arXiv preprint arXiv:2203.05794},
  year={2022}
}

@misc{grootendorst2020keybert,
  author       = {Maarten Grootendorst},
  title        = {KeyBERT: Minimal keyword extraction with BERT.},
  year         = 2020,
  publisher    = {Zenodo},
  version      = {v0.3.0},
  doi          = {10.5281/zenodo.4461265},
  url          = {https://doi.org/10.5281/zenodo.4461265}
}

@article{eloundou2023gpts,
  title={Gpts are gpts: An early look at the labor market impact potential of large language models},
  author={Eloundou, Tyna and Manning, Sam and Mishkin, Pamela and Rock, Daniel},
  journal={arXiv preprint arXiv:2303.10130},
  year={2023}
}

@inproceedings{chen2016xgboost,
  title={Xgboost: A scalable tree boosting system},
  author={Chen, Tianqi and Guestrin, Carlos},
  booktitle={Proceedings of the 22nd acm sigkdd international conference on knowledge discovery and data mining},
  pages={785--794},
  year={2016}
}

@article{ke2017lightgbm,
  title={Lightgbm: A highly efficient gradient boosting decision tree},
  author={Ke, Guolin and Meng, Qi and Finley, Thomas and Wang, Taifeng and Chen, Wei and Ma, Weidong and Ye, Qiwei and Liu, Tie-Yan},
  journal={Advances in neural information processing systems},
  volume={30},
  year={2017}
}

@inproceedings{reimers-2019-sentence-bert,
    title = "Sentence-BERT: Sentence Embeddings using Siamese BERT-Networks",
    author = "Reimers, Nils and Gurevych, Iryna",
    booktitle = "Proceedings of the 2019 Conference on Empirical Methods in Natural Language Processing",
    month = "11",
    year = "2019",
    publisher = "Association for Computational Linguistics",
    url = "http://arxiv.org/abs/1908.10084",
}

@inproceedings{ge2020understanding,
  title={Understanding echo chambers in e-commerce recommender systems},
  author={Ge, Yingqiang and Zhao, Shuya and Zhou, Honglu and Pei, Changhua and Sun, Fei and Ou, Wenwu and Zhang, Yongfeng},
  booktitle={Proceedings of the 43rd international ACM SIGIR conference on research and development in information retrieval},
  pages={2261--2270},
  year={2020}
}

@article{yiu2024ai,
  title={AI Exposure and Strategic Positioning on an Online Work Platform},
  author={Yiu, Shun and Seamans, Rob and Raj, Manav and Liu, Ted},
  journal={arXiv preprint arXiv:2403.15262},
  year={2024}
}

@misc{genai-glossary,
  author       = {Civitai Education},
  title        = {Generative AI Glossary},
  year         = {2023},
  url          = {https://education.civitai.com/generative-ai-glossary/}
}

\clearpage
\appendix
\section{A Glossary of GenAI Models.}
\begin{table}[!htb]
\resizebox{.9\linewidth}{!}{%
\begin{tabular}{|L{8em}|L{30em}|}
\toprule
\textbf{Category} & \textbf{Keywords} \\ \midrule
\textbf{ChatGPT} & ChatGPT, GPT3.5, GPT-3.5, GPT35, GPT-4, GPT4, GPT-4o, GPT4o \\ \midrule
\textbf{DALL-E} & DALL-E, DALL·E, DALLE \\ \midrule
\textbf{Copilot} & Copilot, Bing AI \\ \midrule
\textbf{Gemini} & Bard, Gemini \\ \midrule
\textbf{Sora} & Sora \\ \midrule
\textbf{Llama} & Llama \\ \midrule
\textbf{MidJourney} & MidJourney \\ \midrule
\textbf{Stable Diffusion} & Stable Diffusion, StableDiffusion \\ \midrule
\textbf{Claude} & Claude \\ \midrule
\textbf{Other keywords} & AGI, AI Art, AI Audio, AI Image, AI Music, AI Painting, AI Photo, AI Picture, AI Song, AI Video, AI Writting, AnimateDiff, CFG, CLIP, Chatbot, Chat Bot, Checkpoint, Civitai, Cmdr2, CodeFormer, CompVis, Conversational AI, DDIM, DPM, Deforum, Diffusion Model, DreamArtist, DreamBooth, DreamStudio, ESRGAN, GAN, GFPGAN, GenAI, Generative AI, Generative Model, LLM, Language Model, Latent Diffusion, LoRA, NovelAI, OpenAI, PLMS, SadTalker, Tokenizer, Vicuna, Whisper, depth2img, img2img, prompt engineering, prompt tuning, txt2img, txt2video \\ \bottomrule
\end{tabular}%
}
\caption{A glossary of ChatGPT and GenAI.}
\label{tab:glossary}
\end{table}

\section{Hyperparameter of Experimented Classifiers.}
\begin{table}[!htb]
\resizebox{.8\linewidth}{!}{%
\begin{tabular}{|L{5.5em}|L{27.5em}|}
\toprule
\textbf{Classifier} & \textbf{Python package (Parameters)} \\
\midrule
SVM & \texttt{sklearn} (Default parameters)
\\ \midrule
RF & \texttt{sklearn} (max\_depth: 10, random\_state: 42)
\\ \midrule
LR & \texttt{statsmodels} (Default parameters)
\\ \midrule
XGBoost & \texttt{xgboost} (objective: ``binary:logistic'', max\_depth: 10, learning\_rate: 0.1, num\_boost\_round: 100)
\\ \midrule
LightGBM & \texttt{lightgbm} (objective: ``binary'', boosting\_type: ``gbdt'', learning\_rate: 0.1, num\_boost\_round: 100) 
\\
\bottomrule
\end{tabular}%
}
\caption{A summary of experimented classifiers and corresponding settings.}
\label{tab:models}
\end{table}

\section{Industries of GenAI-related Job Posts}
\begin{figure}[!htb]
  \centering
  \includegraphics[width=.9\linewidth]{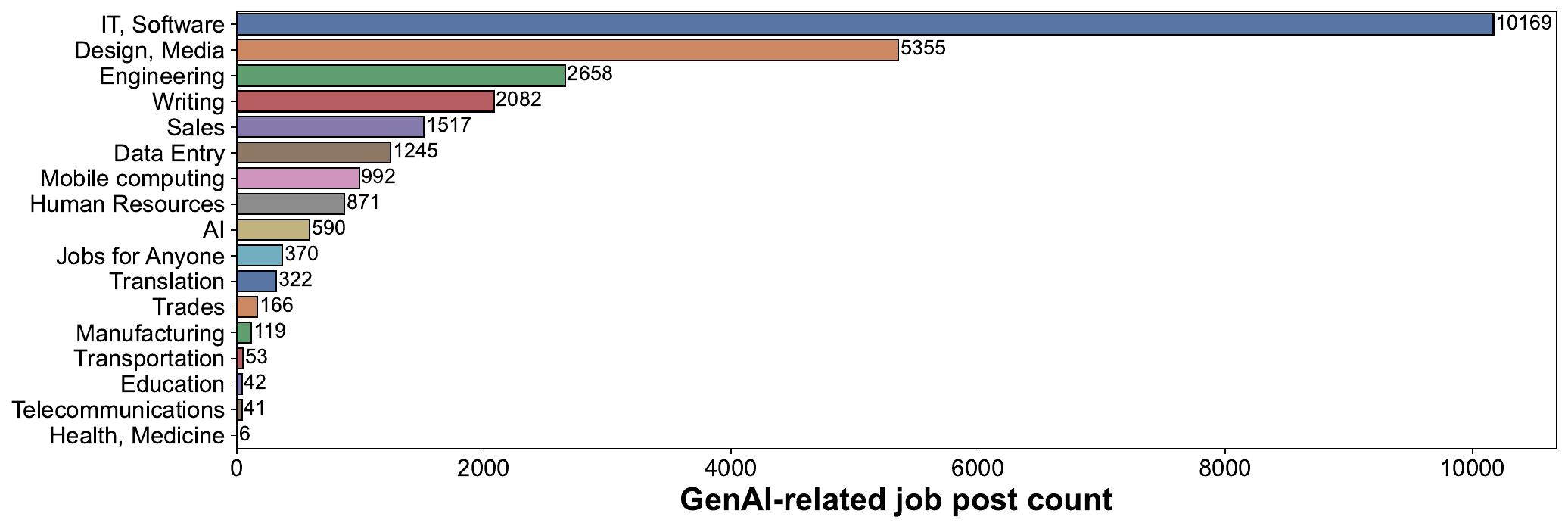}
  \caption{Top industries of GenAI-related job posts.}
  \label{fig:genai_industry}
\end{figure}

\section{Top GenAI Models of GenAI-related Job Posts}
\begin{figure}[!htb]
  \centering
  \includegraphics[width=.9\linewidth]{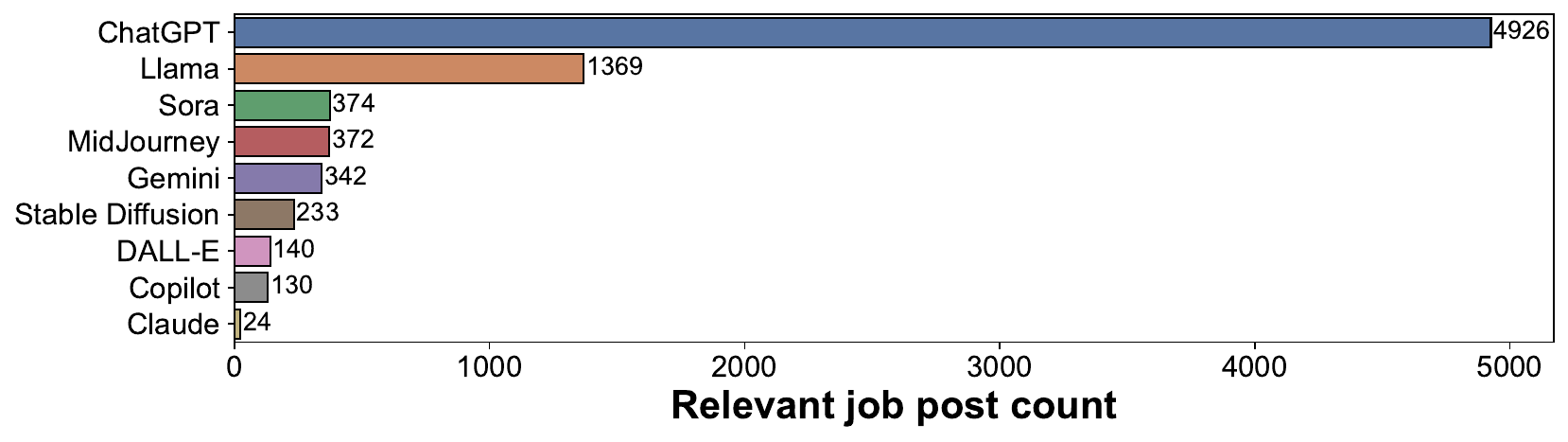}
  \caption{Distribution of job post count for diverse GenAI.}
  \label{fig:genai_post_cnt}
\end{figure}

\newpage

\section{Visualization and Statistics for Topical analysis.}
\begin{figure}[!htb]
  \centering
  \includegraphics[width=.9\linewidth]{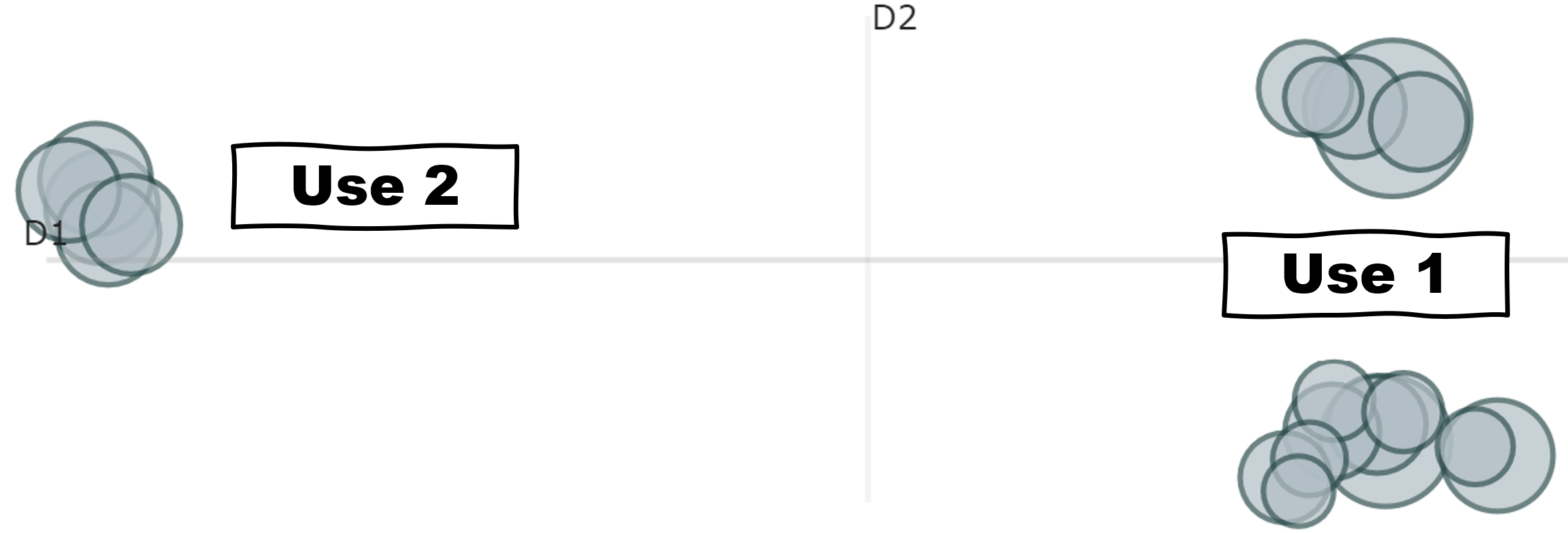}
  \caption{An intertopic distance map generated by BerTopic.
  The distribution of topical clusters indicates the existence of two main use of ChatGPT.
  }
  \label{fig:topic_distance}
\end{figure}

\begin{figure}[!htb]
  \centering
  \includegraphics[width=.8\linewidth]{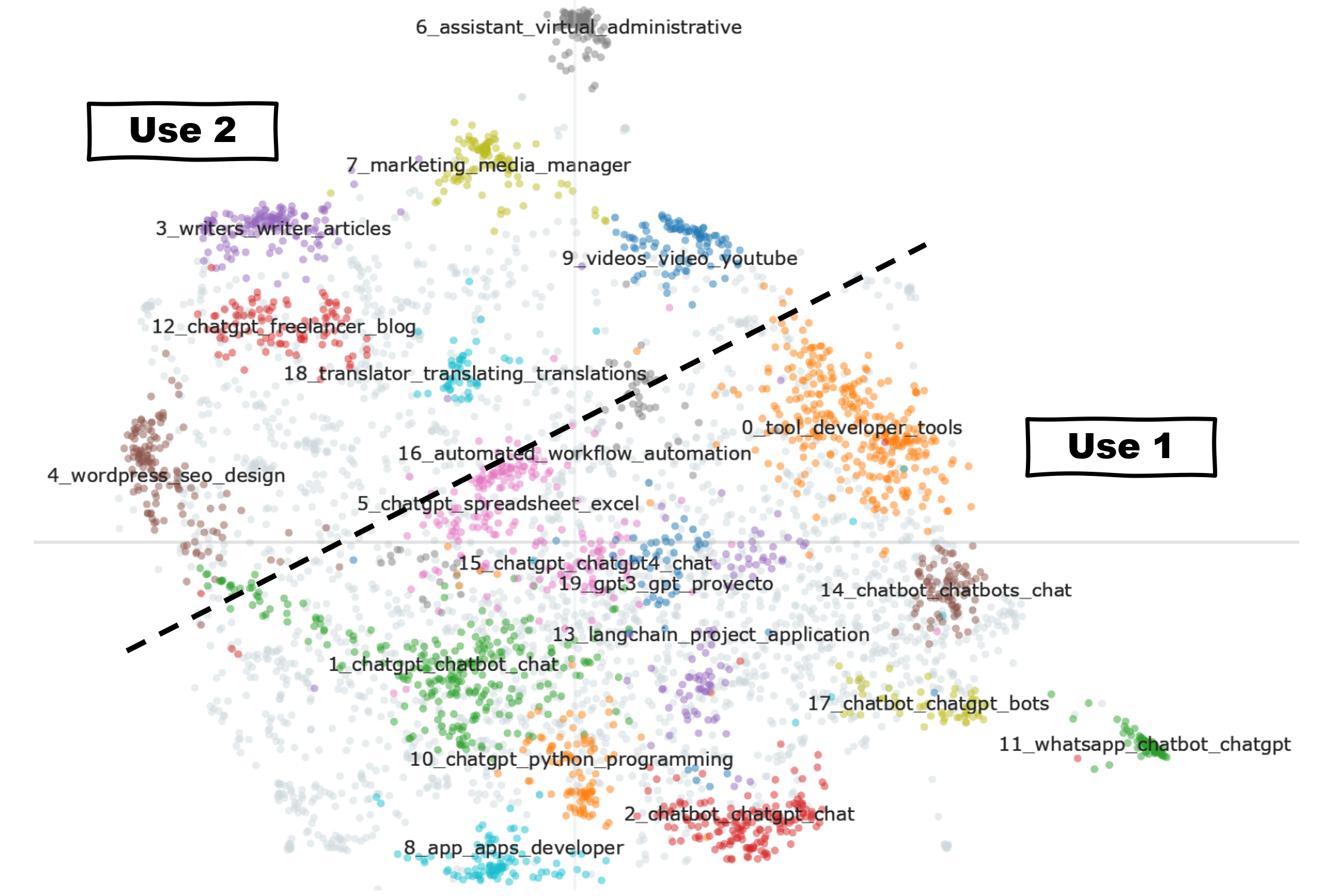}
  \caption{Visualization of job posts' embeddings and topical clusters. The textual labels of topical clusters denote the indexes and representative words of corresponding topics.}
  \label{fig:topic_vis}
\end{figure}

\begin{figure}[!htb]
  \centering
  \includegraphics[width=.8\linewidth]{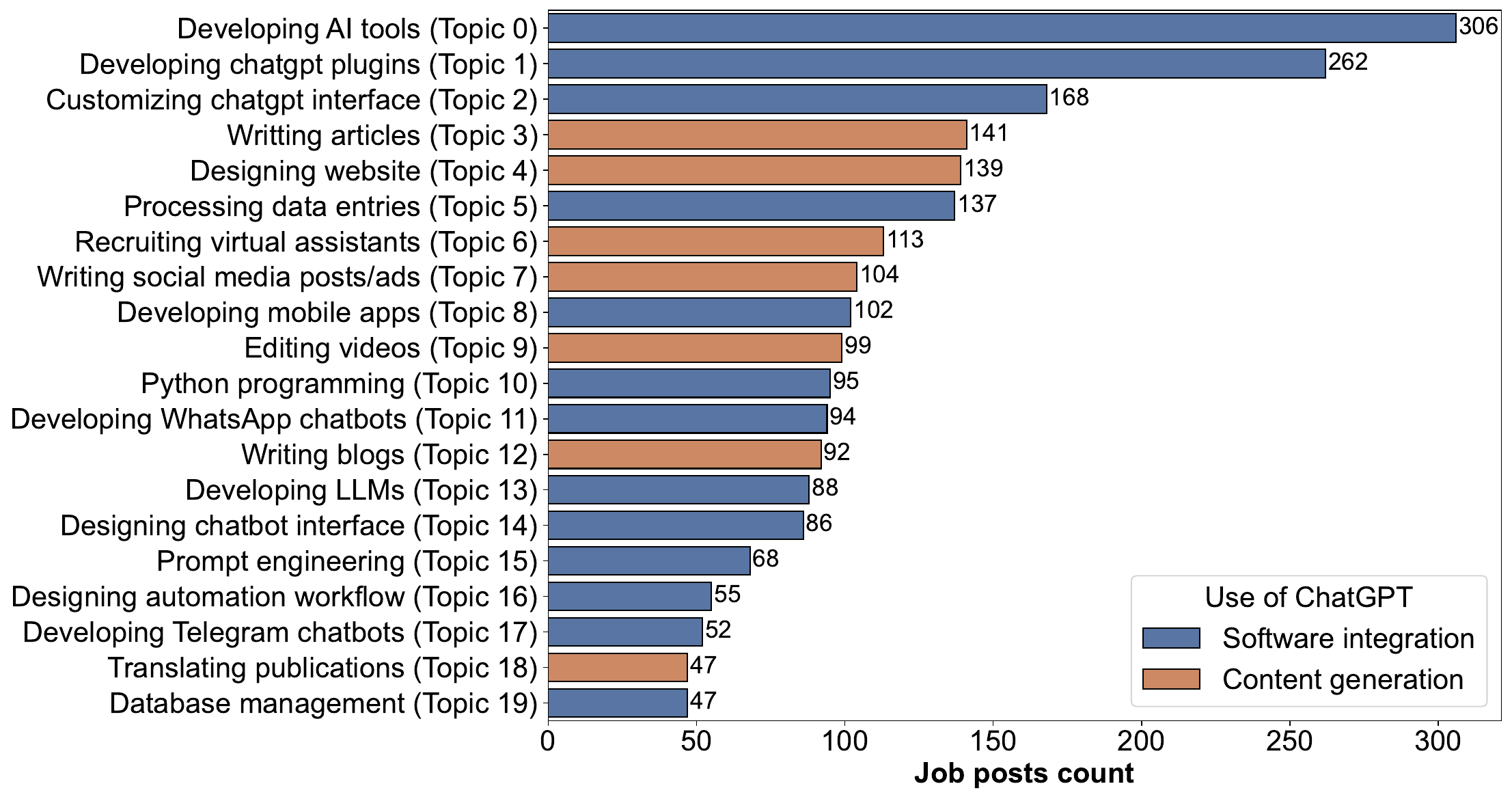}
  \caption{Frequency of the top 20 most popular topics in GPT-related job posts.}
  \label{fig:topic_bars}
\end{figure}

\clearpage
\section{Job Content Related to ChatGPT}
\label{appd:job_content}
\begin{itemize} [leftmargin=*]
    \item \textbf{Use 1: Integrating ChatGPT into software development.} 
    A common goal of employers is to ask for workers who can integrate ChatGPT into specific software and web applications. This target pertains to 13 topics, covering 1,560 (62.02\%) GPT-related job posts from non-outlier topics. Employers mainly recruit people to integrate ChatGPT to 
    \one~develop AI-powered tools or platforms (Topic 0: ``developer, artificial, application, AI, platform, tool'', 10.47\% GPT-related job posts from non-outlier topics);
    \two~create AI chatbot plugins/interfaces (Topic 1: ``chatbot, chatgpt, chat, plugin'', 8.97\%; and Topic 2: ``chatbot, chatgpt, interface, application'', 3.61\%);
    and 
    \three~perform data entry or processing (Topic 5: ``chatgpt, spreadsheet, excel, task, sheet, data'', 4.69\%).
    
    \item \textbf{Use 2: Reproducing human-generated content.} 
    Another common goal of employers is to recruit workers who can use ChatGPT to reproduce human-generated content. This theme contains 7 topics, covering 735 (25.15\%) analyzed job posts. Workers are mainly asked to use ChatGPT to
    \one~reproduce articles (Topic 3: ``writer, writing, article, content, editing'', 4.83\%);
    \two~design websites (Topic 4: ``wordpress, seo, design, website, web'', 4.76\%);
    and
    \three~generate social media posts/ads (Topic 7: ``marketing, media, manager, influencer, promote, communication, ads'', 3.56\%), and weblogs (Topic 12: ``chatgpt, blog, copywriter, content, writer, publish'', 3.15\%). 
    Interestingly, some employers expose their wish to hire virtual assistants capable of constantly customizing GPT-generated content, in order to facilitate administrative work (Topic 6: ``assistant, virtual, administrative, manage, tasks'', 3.87\%), edit videos (Topic 9: ``video, youtube, editing, editor, creator, channel'', 3.39\%), and translate publications (Topic 19: ``translator, translations, traducción, language, idioma'', 1.61\%).
    
\end{itemize}

\newpage

\section{Earliest periods after ChatGPT's release when Classifiers Achieves a Stable Classification.}
\begin{table}[!htb]
\centering
\resizebox{.9\linewidth}{!}{%
\begin{tabular}{|L{6em}|C{10em}|C{10em}|}
\toprule
\textbf{Classifier} & \textbf{Target: relevant} & \textbf{Target: irrelevant} \\ \midrule
{SVM} & +360 & +210 \\ \midrule
{RF} & +330 & +240 \\ \midrule
{LR} & +390 & +300 \\\midrule
{XGBoost} & +330 & +210 \\ \midrule
{LightGBM} & +330 & +210 \\ \bottomrule
\end{tabular}%
}
\caption{The earliest period after ChatGPT's release when classifiers achieves a stable classification for job skills relevant/irrelevant to ChatGPT market. ``+N'' denotes N days after ChatGPT's release.}
\label{tab:early_market}
\end{table}
\end{document}